\pdfoutput=1

\documentclass[11pt]{article}

\usepackage{ACL2023}

\usepackage{times}
\usepackage{latexsym}
\usepackage{censor}

\usepackage[T1]{fontenc}

\usepackage[utf8]{inputenc}
\usepackage{graphicx}
\graphicspath{ {./images/} }
\usepackage{booktabs}
\newcommand{\ra}[1]{\renewcommand{\arraystretch}{#1}}
\usepackage{tabularx}
\usepackage{multirow}
\usepackage{amssymb}
\usepackage{pifont}
\newcommand{\cmark}{\ding{51}}
\usepackage{hyperref}
\usepackage{dblfloatfix}

\usepackage{microtype}

\usepackage{inconsolata}

\title{Legal NLP Meets MiCAR:\\Advancing the Analysis of Crypto White Papers}

\author{Carolina Camassa \\
  Bank of Italy  \\
  \texttt{carolina.camassa@bancaditalia.it} \\}

\begin{document}
\maketitle
\begin{abstract}
In the rapidly evolving field of crypto assets, white papers are essential documents for investor guidance, and are now subject to unprecedented content requirements under the European Union's Markets in Crypto-Assets Regulation (MiCAR). Natural Language Processing (NLP) can serve as a powerful tool for both analyzing these documents and assisting in regulatory compliance. This paper delivers two contributions to the topic. First, we survey existing applications of textual analysis to unregulated crypto asset white papers, uncovering a research gap that could be bridged with interdisciplinary collaboration. We then conduct an analysis of the changes introduced by MiCAR, highlighting the opportunities and challenges of integrating NLP within the new regulatory framework. The findings set the stage for further research, with the potential to benefit regulators, crypto asset issuers, and investors.

\end{abstract}


\section{Introduction}\label{introd}
White papers are the cornerstone of information disclosure for Initial Coin Offerings (ICOs), a modern fundraising method that sells tokens to a diverse group of investors \citep{Fisch2019}. These fundraisings represent an alternative method for entrepreneurial finance, with similarities to IPOs, venture capital, and presale crowdfunding \cite{howell2020initial}. \\
When a new ICO is announced, it is usually accompanied by the \textit {white paper}, a text document that contains information for the investors and bears some similarities, at least in intent, with an IPO prospectus. However, unlike their regulated counterparts, white papers have so far operated in a largely unregulated landscape. \newline
Thanks to this regulatory vacuum, the type of content varies widely, but it usually includes \cite{bourveau2022role, zhang2019readability} a description of the service offered and its value proposition, the composition of the project team, financial details about the offer, and explanations of the technologies underlying the project.
Due to their public accessibility and richness in data, white papers have been a frequent subject of studies aimed at extracting market insights through textual analysis (\citet{Fisch2019, unpackingblackbox, expertsorcharlatans} among others).

The current ICO market is characterized by a pronounced level of \textit{information asymmetry} between crypto issuers and investors \cite{bourveau2022role,chen2019information}. These circumstances, easily explained by the lack of regulation, led to investigation into the prevalence of fraud and scams among ICOs \cite{liebau2019crypto,karimov2021identification}. It is not uncommon for white papers to be a vehicle for misinformation, sometimes resorting to imprecise claims and exaggerated language \cite{momtaz2021entrepreneurial} to lure investors.
%
%

\begin{figure*}[b!]
\centering
  \includegraphics[width=\textwidth]{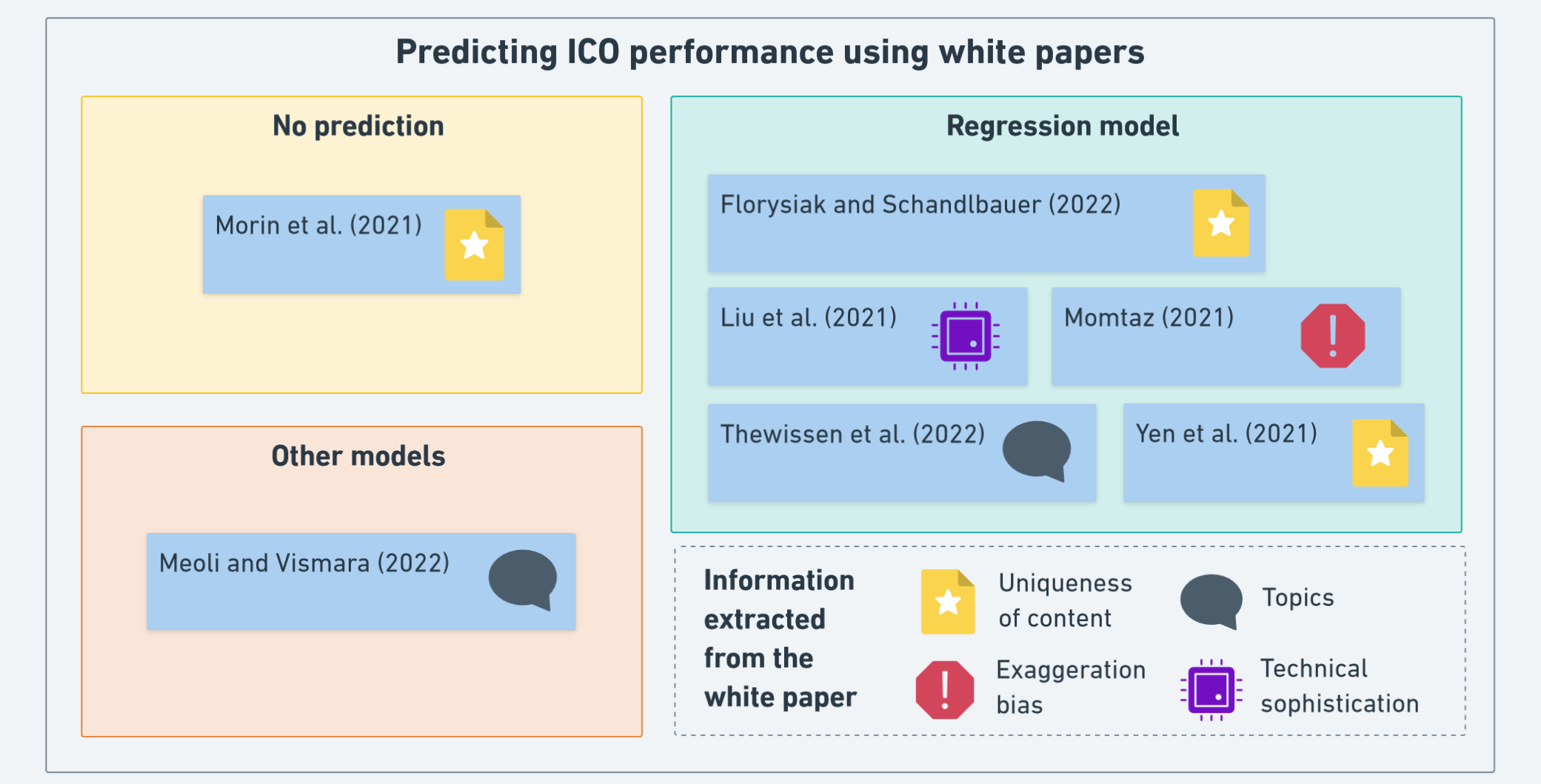}
  \caption{A shared research objective in existing studies that apply NLP to crypto asset white papers is predicting the degree of success of an ICO. The figure shows the experiments surveyed in section \ref{sec:relatedwp}, grouped according to the type of prediction model used and the variables extracted from the text. }
  \label{figure:whitepapers-survey}
\end{figure*}


\subsection{Introduction to the MiCA Regulation}

In this context of fragmented or absent regulation of the crypto assets market, the European Union has been making an effort to unify regulation in its member states on this aspect. The vehicle for this unification is the \textit{Regulation on Markets in Crypto Assets} (abbreviated as MiCAR).
The text of the regulation was published on the Official Journal of the European Union on June 9th, 2023 \cite{micar} \footnote{OJ L 150, 9.6.2023, p. 40–205}, and the norms will become fully applicable by December 30th, 2024. Its entry into force will provide clarity and protection for European citizens, but it will also require considerable effort on the part of issuers \cite{florysiak2022utility}.

The MiCAR aims to regulate crypto asset public offerings---commonly known as ICOs---the admission of those crypto assets to trading on platforms, and additional related services to customers. \\
One key focus is the standardization of white paper content; the regulation mandates the inclusion of specific elements, such as details about the token offering, the associated risks, and team composition, with an emphasis on textual clarity and readability. 

These new rules are engineered not just to to enable European citizens to make informed investment decisions, but also to assist national authorities in monitoring crypto asset offerings.

\subsection{Motivation and main contributions} 
The goal of this study is to overview current Natural Language Processing applications to the analysis of crypto asset white papers, and identify how new applications can assist stakeholders in adapting to the upcoming MiCA regulation. \\
Given that much of the existing research comes from economics and finance, this work aims to bridge the domain gap, enabling future Computer Science (CS) researchers to leverage all relevant work on the topic. It also provides a starting point for researchers and practitioners in the legal NLP domain to adapt existing methods to this new topic.\\
The organization of this work is as follows:
\begin{itemize}
    \item  A survey of NLP applications to crypto asset white papers and Initial Public Offering (IPO) prospectuses is provided in Section \ref{related}. 
    \item Section \ref{sec:micar} goes into detail on how the new MiCA regulation will impact the structure and  content of white papers. 
    \item Finally, Section \ref{NLPopportunities} highlights the challenges and opportunities of harnessing NLP techniques to aid all stakeholders involved in the MiCAR compliance process.

\end{itemize}



\begin{table*}[ht!]  

\ra{1.2}
\begin{tabularx}{\textwidth}{
  >{\hsize=2\hsize}X
  >{\hsize=0.8\hsize}X
  >{\hsize=0.8\hsize}X
  >{\hsize=0.8\hsize}X
  >{\hsize=0.8\hsize}X
  >{\hsize=0.8\hsize}X
  @{}}
\toprule
Study         & Dataset size         & Sentiment analysis & Topic\newline modeling & Document \newline similarity & NER 
\\
\midrule
\citet{expertsorcharlatans} & 2,665 &  &   &  \cmark &  \cmark \\ 
\citet{liu2021technology} & 1,629 &  \cmark & \cmark &  &  \\ 
\citet{meoli2022machine} & 451 & \cmark & \cmark &  & \cmark  \\ 
\citet{momtaz2021entrepreneurial} &  495 & \cmark &  &   &\\
\citet{morin2021detecting} & 3,299  &   &   &  \cmark \\ 
\citet{unpackingblackbox} &  5,210 &  \cmark  &  \cmark   &  & \cmark  \\ 
\citet{yen2021different} & 1,064 &   &   &  \cmark & \\ 
\bottomrule 
\end{tabularx}
\captionof{table}{Overview of existing literature applying NLP techniques to crypto asset white papers, along with the dataset size and the techniques used. Most works come from the economics and finance domains.}\label{table:whitepapers-survey}
\end{table*}
\section{Related work: NLP in ICO White Papers and IPO Prospectuses}\label{related}

\subsection{NLP and ICO white papers}\label{sec:relatedwp}
A survey of the field shows that most existing studies on crypto asset white papers are published in economics and finance venues, with little overlap with CS research.

A recurring research goal is using information extracted from the content of white papers to make predictions about the likelihood of success of an ICO. In this context, the type of signal extracted can vary widely, as do the predictive models used. Some studies on the topic relies exclusively on manual text analysis \cite{bourveau2022role, thewissen2023linguistic,Fisch2019}; this review focuses instead on those studies that incorporate at least one NLP method, setting dictionary-based approaches as the minimum threshold for inclusion\footnote{Excluded work typically relies on page length \cite{samieifar2021read,bourveau2018initial} or readability metrics such as the Gunning-Fog index or the Flesch Reading Ease score \cite{zhang2019readability}.}. \\
Figure \ref{figure:whitepapers-survey} shows a taxonomy of the studies grouped according to the variables of interest and the type of model, and Table \ref{table:whitepapers-survey} lists the dataset size and NLP tasks for each.

\citet{momtaz2021entrepreneurial} uses a dictionary-based feature extraction method first proposed by \citet{hu2004mining} to capture the \textit{exaggeration bias} in white papers. Momtaz argues that, in the absence of regulation, issuers of crypto asset tokens may have an incentive to positively exaggerate the quality of their venture to attract investors.
The analysis confirms that there is a pervasive positive bias in the content of white papers, and that ICOs with this characteristic raise more funds in less time. 

A common strategy is to use variations of the Latent Dirichlet Allocation (LDA) algorithm \cite{blei2003latent} to extract topic information from the white papers. The output of the model can be used, for example, to quantify the amount of technical information included in the text \citep{liu2021technology}. This variable shows that crypto assets with high \textit{technology indexes} are more likely to have positive long-run performances of the associated ICOs.
\citet{unpackingblackbox} use \textsc{sentLDA} \citep{sentLDA}, a sentence-level version of LDA, to extract 30 topics from an extensive collection of white papers. The topics generated by the algorithm are evaluated and manually labeled by the authors, and the topic assignment is then used as a regression variable. The results of the regression show that the topics contained in a white paper can partially explain the success of ICOs.

Given the large amount of new ICOs published each year, some research has focused on looking for similarities among published white papers. 
Using a term-frequency approach initially developed for IPO prospectuses by \citet{hanley2010information}, \citet{expertsorcharlatans} and \citet{yen2021different} assess the informational value of these documents. Here, a white paper is considered more informative if it introduces new or additional information not found in similar papers. The outcomes indicate a correlation between high informational content, or uniqueness, and success metrics such as fundraising and post-ICO market values. Separately, \citet{morin2021detecting} investigate text similarity across white papers using three metrics: TF-IDF, cosine similarity, and pairwise similarity. Their findings reveal that 19\% of ICO white papers exhibit high similarity to previously published ones.\\
Finally, \citet{meoli2022machine} combine sentiment analysis, NER, topic modeling, sentiment analysis and document structure analysis to extract useful features from the text. The features are then used to train a set of binary classifiers, with different machine learning approaches, to predict whether an ICO was successful or not. The best machine learning model outperforms the benchmark, a traditional econometric forecasting model. A feature importance analysis confirms the relevance of the information extracted from the text, especially the document structure and the sentiment score.
\begin{table*}[!th]
\centering

\ra{1.3}
\begin{tabularx}{\textwidth}{lX}
\toprule
Part & Description \\ 
\midrule
A & Information about the offeror or the person seeking admission to trading \\
B & Information about the issuer, if different from the offeror or person seeking admission to trading \\
C & Information about the operator of the trading platform  \\
D & Information about the crypto asset project \\
E & Information about the offer to the public of crypto assets or their admission to trading \\
F & Information about the crypto assets \\
G & Information on the rights and obligations attached to the crypto assets \\
H & Information on the underlying technology \\
I & Information on the risks \\ \bottomrule
\end{tabularx}
\caption{Content requirement categories for crypto assets ``other than''---one of the three types of assets regulated by MiCAR.}
\label{table:requirements-list}
\end{table*}
\subsection{NLP and IPO prospectuses}\label{sec:relatedIPO}

Textual analysis methods have also been used to extract information from the contents of IPO prospectuses. A prospectus is a document required by national financial regulatory authorities to present an investment offering---such as stocks, bonds, and mutual funds---to the public, or to obtain admission to trading on a regulated market. It contains all the details of the financial offer and must inform investors of the risks involved with investing. These features make the documents partially similar, at least in intent, to crypto asset white papers.

\citet{tao2018analysing} train a custom embedding model using the \textsc{word2vec} algorithm \cite{mikolov2013distributed}, to find and analyze sentences in the text that represent ``forward looking statements'' (FLS). These statements are meant to provide useful information about the company’s future performance. \citet{li2018incorporating} apply a keyword-based approach to the same task. \\
\citet{sharpe-decker-2022-prospectus} perform sentiment analysis to look for the effect of text sentiment on the probability of an IPO being withdrawn.\\
Finally, \citet{hanley2010information} devise a method---previously mentioned in \ref{sec:relatedwp}---to separate the information content of a prospectus into \textit{standard} and \textit{informative} components. The components are derived from term frequency data and they are used to quantify how much of the content is also found in documents published in the same time period---or in the same industry---and how much of it is unique to a specific document.
\subsection{Conclusions from the survey}

While there seems to be a healthy amount of research interest in the topic, most of the existing literature covers a very limited array of NLP techniques and tasks. As an hypothesis, this might be a result of the lack of intersection between this particular domain of economics and finance research and CS research.
\newline More advanced NLP methodologies have succesfully been applied in various regulatory and legal settings. The following section contains examples of such works and argues for the transferability of these computational approaches to the analysis and regulation of future crypto asset white papers.

\section{How MiCA changes the regulatory landscape}\label{sec:micar}
The MiCA regulation is broad and covers three categories of crypto assets: \textit{asset-referenced tokens}, \textit{e-money tokens}, and \textit{crypto assets other than}\footnote{For brevity; the full definition in the text is ``crypto assets other than asset-referenced tokens or e-money tokens''.}.
The latter category also includes \textit{utility tokens}, a specific type of crypto asset that has no financial purposes and is only intended to provide access to a good or a service supplied by its issuer.
Some types of crypto assets, such as \textit{Non-Fungible Tokens} (NFTs), are not covered by the regulation. 
\newline
For each crypto asset category , the regulation contains norms regarding the content and issuance of the white paper, the conduct of business, in addition to organizational and financial requirements.
\par
In alignment with the objective of the study, the rest of the work will focus only on the norms for white paper content, and in particular those valid for the crypto ``other than'' category. While the requirements vary across the three asset categories, the differences are marginal in the scope of the analysis. Another reason supporting this choice is the fact that the ``other than'' category has the largest set of requirements for white papers---see Appendix \ref{sec:appendixA}.

\paragraph{\textbf{Minimum content requirements.}}
Each whitepaper must satisfy a comprehensive list of minimum content requirements\footnote{As outlined in Annexes I, II, and III of the regulation.} that are organized around macro-areas, listed in Table \ref{table:requirements-list} in the case of ``other than'' crypto assets. A comparison with the other two categories is provided in Appendix \ref{sec:appendixA}.

\begin{table*}[hb]
    \centering
    \ra{1.3}
    \begin{tabularx}{\textwidth}{lXX}
        \toprule
        Part & Disclosure item & White paper sample \\
        \midrule
        \multirow{2}{*}{E} & E5: the total number of crypto assets to be offered to the public or admitted to trading. & A total of \textcolor{ForestGreen}{200 million tokens} is put into circulation through a private sale and a public ICO. \\
         \cmidrule{2-2}
        
        & E14: Information about technical requirements that the purchaser is required to fulfil to hold the crypto assets.  & To engage with the \censor{XXX} protocol, individuals \textcolor{ForestGreen}{must first download \censor{XXX} App.} \\
        \hline
        \multirow{2}{*}{H} & H1: The consensus mechanism, where applicable. & The \textcolor{ForestGreen}{IBFT PoA} is the default consensus mechanism on \censor{Galo} Blockchain. \\
       \cmidrule{2-2}
        & H5: Information on the audit outcome of the technology used, if such an audit was conducted. & \textcolor{ForestGreen}{Two separate security assessments} were conducted over a period of several months and conducted by the audit firms \censor{audit1} and \censor{audit21}. \\
        \hline
        \multirow{1}{*}{I} & I4: A description of the risks associated with project implementation. & (...), no security measures can provide absolute protection against \textcolor{ForestGreen}{unauthorized access and data breaches.} \\
        
         \bottomrule
    \end{tabularx}
    \caption{Examples of the minimum content requirements from MiCAR, accompanied by redacted excerpts taken from existing, unregulated, white papers. The first column shows the requirement category. All categories and their descriptions are listed in Table \ref{table:requirements-list}.}
    \label{table:examples}
\end{table*}

Table \ref{table:examples} shows some examples of these requirements, accompanied by related excerpts taken from existing (pre-regulation) white papers.
\paragraph{\textbf{Other requirements}}
In addition to minimum content requirements, other MiCAR norms impact white paper content and could have implications for future NLP applications on the topic.
\begin{itemize}
    \item \textbf{Future value of the assets:} issuers of crypto assets are forbidden\footnote{Articles \textit{6.4, 6.5a, 19.3, 19.4a}.} from making claims about the future value of the asset in the white paper, aside from  clarifying that the token might lose its value partially or in full \footnote{Does not apply to \textit{e-money tokens}.}. These kinds of forward-looking statements have previously been analysed in IPO prospectuses---see Section \ref{sec:relatedIPO}.
    \item \textbf{Document language:} according to the regulation\footnote{Articles \textit{6.9}, \textit{19.8}, and \textit{51.8}.},  \textit{``the white paper shall be drawn up in an official language of the home Member State, or in a language customary in the sphere of international finance''}. Using a language other than English could pose some challenges that are examined in Section \ref{NLPopportunities}.
    \item \textbf{Document template:} the MiCA regulation states\footnote{Articles \textit{6.10-6.12}} that the white paper must adhere to the template and form established by the European Securities and Markets Authority (ESMA) and the European Banking Authority (EBA). As of October 2023, no template has been published yet, although consultations on technical requirements are in progress.
\end{itemize}


\section{NLP in the MiCAR Landscape: Opportunities and Challenges}\label{NLPopportunities}

\subsection{What NLP can do: exploring possibilities.}\label{sec:nlp-tasks}
Even if there is currently a gap in NLP research applied to the topic of crypto white papers, other legal domains have benefited from collaboration with CS researchers \cite{zhong-etal-2020-nlp}. 

Tasks involving the interpretation and analysis of legal texts have been a popular subject for the legal NLP community. Current research mostly relies on pre-trained Transformer-based \cite{vaswani2017attention} models, with rising interest in the use of Large Language Models (LLMs) \cite{guha2023legalbench,yu-etal-2023-exploring}.


The rest of the section explores some avenues for applying existing NLP techniques to the MiCA regulatory framework.

\paragraph{\textbf{Predicting the outcome of compliance checks.}} 
Given a text passage from a white paper and a MiCAR requirement, it would be beneficial to have a model capable of correctly and automatically check if the passage complies with the regulation. \newline
There have been attempts to solve this problem for different regulatory domains. For example, \citet{https://doi.org/10.48550/arxiv.2004.03422} attempt to automate the decision of whether an online post is subject to the European Union's Legal Framework against the Expression of Hatred. As part of their approach to classify hate speech from a legal standpoint, they trained \textsc{GBERT} \cite{chan-etal-2020-germans} as a multi-class classifier on a manually annotated hate speech dataset. The model showed poor performance, suggesting that the task might be too complex for the model's capabilities. 
The task could be similarly challenging in the case of MiCAR, making it a good candidate for further research.

\paragraph{\textbf{Matching passages with requirements.}} 
If automating compliance checks proves too difficult---or even undesirable---another useful application could be simply to identify which regulation norms are relevant to each text passage. \\
In a similar task, \citet{ravichander-etal-2019-question} collects a set of annotated questions about the contents of privacy policies, in which the model must determine if the text passage contains an answer to the provided question, and classify the question-passage pair as Relevant or Irrelevant. \\
\citet{Abualhaija_2022} use language models from the \textsc{BERT} family \cite{DBLP:conf/naacl/DevlinCLT19}, finetuned on the task of Question Answering, to assist requirements engineers in finding text passages relevant to their analysis of compliance requirements.

\paragraph{\textbf{Question answering and information retrieval.}} 
Letting users ask questions about the content of a white paper---be it a potential investor trying to understand the project, or a regulator looking for a specific paragraph---is one of the most obvious and potentially useful applications of NLP to this topic. \\
Given the length and complexity of the regulation, stakeholders could also benefit from an efficient way to retrieve norms from the MiCAR text.\\
QA and IR systems in the legal and regulatory domain pose some specific challenges compared to traditional question answering \cite{abdallah2023exploring}. Among others, they usually require highly specialized datasets curated by legal experts, and there is less room for error since an imprecise or factually wrong answer could negatively impact a legal decision.

To mitigate the data problem, the authors of \textsc{LegalBench} \cite{guha2023legalbench} collect over 150 annotated datasets in the legal domain, including QA tasks. Since regulated white papers will be publicly available, the research community could contribute to this effort in the future with a QA dataset inspired by the MiCA regulation.

\paragraph{\textbf{LLMs for reasoning and summarization.}} 
As demonstrated in \textsc{LegalBench}  \cite{guha2023legalbench}, Large Language Models can now achieve good performance on some challenging legal reasoning tasks in a zero-shot setting. While supervised approaches may still outperform these models, there is a clear advantage in using LLMs as they don't require labeled datasets.
Selecting the appropriate instruction and few-shot examples for the prompt can further improve performance \cite{yu-etal-2023-exploring, trautmann2022legal}. 

Aside from reasoning, another promising application of LLMs is summarization. The emergence of training methods that use human preference data \cite{stiennon2020learning} has led to models that can produce high-quality summaries of text, including those in the legal domain \cite{pont2023legal}, without domain-specific training. In the MiCAR context, both the regulation text and the white papers are long and often dense, making summarization a useful tool for regulators and investors interested in quickly analyzing their contents. 



\paragraph{\textbf{Named Entity Recognition.}} Some of the MiCAR requirements entail entity extraction, from the traditional kind (e.g., the list of persons involved in managing the project) to more domain-specific types of entities such as Legal Norm and Organization. 
Named entities are also useful for document organization and search. 
However, NER algorithms are particularly sensible to domain shifts: When general-purpose NER algorithms are used in texts from a narrow domain, there is often a performance degradation. To work around this limitation, \citet{au-etal-2022-e} and \citet{smadu-etal-2022-legal} train entity recognition algorithms on annotated legal datasets in different languages.
Given that crypto asset white papers are also financial documents, NER models trained on data from the financial domain might also be an option \cite{salinas-alvarado-etal-2015-domain,francis2019transfer}.

\subsection{Challenges}
NLP can simplify the understanding of crypto asset white papers, aid regulators in MiCAR compliance, and reduce the effort required for issuers to generate compliant documents. It can also benefit investors by enabling them to make more informed decisions, if they are given the tools to better assimilate the contents of white papers.\newline
However, there are also potential obstacles due to the narrow scope of application and the complexity of the subject matter. 

\paragraph{\textbf{Adapting to the legal and regulatory domain.}}

As previously mentioned, performance degradation can occur when using general-purpose language models on narrow domains. This is not limited to entity recognition, but also impacts other tasks. One way of diminishing this effect could be making use of the models released by the legal NLP research community, such as \textsc{LEGAL-BERT} \cite{chalkidis-etal-2020-legal} and \textsc{PolBERT}, a subsequent model trained on the Pile of Law \cite{hendersonkrass2022pileoflaw} dataset. \newline
Since white papers contain terms and language specific to the financial domain, models like \textsc{FinBERT} \cite{araci2019finbert} and \textsc{FinGPT} \cite{2306.06031} could also be appropriate. However, the world of crypto assets brings even more lexical issues with its acronyms and neologisms\footnote{The \href{https://www.gemini.com/it-it/cryptopedia/glossary}{Cryptopedia glossary} from Gemini contains over 1000 terms unique to the crypto community.}.

\paragraph{\textbf{Domain shift between white papers and regulation.}}

In the past, the issuers of crypto asset white papers have used very different language compared to the legal jargon found in regulatory texts. 
This difference might make it harder to find semantic similarities between white paper passages and the relevant regulation articles---one of the possible applications mentioned in Section \ref{sec:nlp-tasks}. \\
\citet{keymanesh2021privacy} encounter a similar issue in developing a QA system that answers citizen queries about privacy policies, and they partially overcome it by using paraphrasing techniques for query expansion.

\paragraph{\textbf{Language.}} The regulation allows using either the local language of the state the crypto asset will be issued in, or a language commonly used in finance. Although we can expect most papers to be in English, the usage of other languages might introduce difficulties, especially in combination with the aforementioned domain shift. There is ongoing research \cite{niklaus2023multilegalpile, chalkidis-etal-2023-lexfiles} that aims to make available language models and NLP datasets that support multiple languages in the legal domain.

\paragraph{\textbf{Document structure and length.}} 
In the largest study encountered \cite{unpackingblackbox}, with a sample of 5210 white papers, the median document length is 30 pages, and the maximum is 167. As the number of required disclosure items rises with the introduction of MiCAR, we might see these numbers increasing.

Handling long documents in NLP presents unique challenges. Both traditional sequence models and transformer-based models struggle, with most \textsc{BERT}-based models having a maximum sequence length of 512 tokens.
Even among recent LLMs, few exceed the limit of 4,096 tokens\footnote{A limit which includes both input and output tokens.}.
To get around this problem, \citet{mamakas-etal-2022-processing} modify \textsc{Longformer} \cite{beltagy2020longformer} and \textsc{Legal-BERT} \cite{chalkidis-etal-2020-legal} to handle texts up to 8,192 tokens.

Until the regulation templates for white papers are released, it cannot be determined whether they will be required to adhere to a provided document structure.
For documents that are not pre-segmented, \citet{aumiller2021structural} propose a segmentation system for legal documents that uses topic modeling to split a given text into semantically coherent spans for downstream applications.

\subsection{Implications for stakeholders}
\textbf{Regulators.}
NLP could help regulators by partially automating compliance checks, reducing the administrative burden on competent authorities. It can also improve the speed and accuracy of oversight with ad-hoc tools for domain experts, enabling them to respond more rapidly to market changes and potential infringements. \\
\textbf{Issuers of crypto assets.}
For issuers, NLP-based tools can assist in drafting white papers that are compliant with MiCAR guidelines from the start. This can reduce the time and costs associated with legal consultations and revisions, making it easier to enter and operate within the European market. \\
\textbf{Potential investors.}
From an investor's perspective, NLP-generated analyses of white papers can provide a more transparent and easily interpretable data source, democratizing access to investment information about the crypto asset market.\\


\section{Conclusions}
Crypto asset white papers are valuable data sources for NLP practitioners due to their public availability and rich informational content. The entry into force of the MiCAR opens new avenues for NLP applications to assist various stakeholders, including regulators, issuers, and potential investors, in navigating compliance and regulatory oversight.

Our survey of existing literature on the textual analysis of white papers revealed an active but siloed field that could benefit significantly from interdisciplinary collaboration between computer science, finance, and legal experts. Existing work in the legal NLP field could serve as a foundation for developing algorithms and models tailored to this new regulatory landscape.

Researchers have a clear opportunity to contribute to this emerging area, with the potential not only to streamline regulatory processes but also to create a more transparent and accountable crypto asset ecosystem.

\section*{Limitations}
The work inevitably contains some speculative elements, given that, while the MiCAR text is public, no regulated white papers are available yet, and no regulatory workflows are yet finalized or known. \\
The new white papers will be expected to follow a template designed by the ESMA, which has not been released yet. Therefore, the analysis of possible NLP applications, opportunities, and challenges is partially grounded in the structures and contents of existing, unregulated white papers, and it may not be entirely applicable to regulated ones.
\section*{Acknowledgments}
The author would like to thank Marco Benedetti and Andrea Gentili for their valuable feedback.
Special thanks are also extended to Claudia Biancotti, Daniele Natalizi, and Mattia Suardi for their insights into clarifying various aspects of the MiCAR regulation.

\bibliographystyle{acl_natbib}
\bibliography{anthology,custom}

\begin{thebibliography}{54}
\expandafter\ifx\csname natexlab\endcsname\relax\def\natexlab#1{#1}\fi

\bibitem[{150(2023)}]{micar}
OJ~L 150. 2023.
\newblock Regulation (eu) 2023/1114 of the european parliament and of the council of 31 may 2023 on markets in crypto-assets.
\newblock \emph{Official Journal of the European Union}, 150:40--205.

\bibitem[{Abdallah et~al.(2023)Abdallah, Piryani, and Jatowt}]{abdallah2023exploring}
Abdelrahman Abdallah, Bhawna Piryani, and Adam Jatowt. 2023.
\newblock Exploring the state of the art in legal qa systems.
\newblock \emph{arXiv preprint arXiv:2304.06623}.

\bibitem[{Abualhaija et~al.(2022)Abualhaija, Arora, Sleimi, and Briand}]{Abualhaija_2022}
Sallam Abualhaija, Chetan Arora, Amin Sleimi, and Lionel~C. Briand. 2022.
\newblock \href {https://doi.org/10.1109/re54965.2022.00011} {Automated question answering for improved understanding of compliance requirements: A multi-document study}.
\newblock In \emph{2022 {IEEE} 30th International Requirements Engineering Conference ({RE})}. {IEEE}.

\bibitem[{Araci(2019)}]{araci2019finbert}
Dogu Araci. 2019.
\newblock Finbert: Financial sentiment analysis with pre-trained language models.
\newblock \emph{arXiv preprint arXiv:1908.10063}.

\bibitem[{Au et~al.(2022)Au, Lampos, and Cox}]{au-etal-2022-e}
Ting Wai~Terence Au, Vasileios Lampos, and Ingemar Cox. 2022.
\newblock \href {https://doi.org/10.18653/v1/2022.nllp-1.22} {{E}-{NER} {---} an annotated named entity recognition corpus of legal text}.
\newblock In \emph{Proceedings of the Natural Legal Language Processing Workshop 2022}, pages 246--255, Abu Dhabi, United Arab Emirates (Hybrid). Association for Computational Linguistics.

\bibitem[{Aumiller et~al.(2021)Aumiller, Almasian, Lackner, and Gertz}]{aumiller2021structural}
Dennis Aumiller, Satya Almasian, Sebastian Lackner, and Michael Gertz. 2021.
\newblock Structural text segmentation of legal documents.
\newblock In \emph{Proceedings of the Eighteenth International Conference on Artificial Intelligence and Law}, pages 2--11.

\bibitem[{Bao and Datta(2014)}]{sentLDA}
Yang Bao and Anindya Datta. 2014.
\newblock \href {http://www.jstor.org/stable/42919610} {Simultaneously discovering and quantifying risk types from textual risk disclosures}.
\newblock \emph{Management Science}, 60(6):1371--1391.

\bibitem[{Beltagy et~al.(2020)Beltagy, Peters, and Cohan}]{beltagy2020longformer}
Iz~Beltagy, Matthew~E Peters, and Arman Cohan. 2020.
\newblock Longformer: The long-document transformer.
\newblock \emph{arXiv preprint arXiv:2004.05150}.

\bibitem[{Blei et~al.(2003)Blei, Ng, and Jordan}]{blei2003latent}
David~M Blei, Andrew~Y Ng, and Michael~I Jordan. 2003.
\newblock Latent dirichlet allocation.
\newblock \emph{Journal of machine Learning research}, 3(Jan):993--1022.

\bibitem[{Bourveau et~al.(2018)Bourveau, De~George, Ellahie, and Macciocchi}]{bourveau2018initial}
Thomas Bourveau, Emmanuel~T De~George, Atif Ellahie, and Daniele Macciocchi. 2018.
\newblock Initial coin offerings: Early evidence on the role of disclosure in the unregulated crypto market.
\newblock \emph{Available at SSRN}, 3193392.

\bibitem[{Bourveau et~al.(2022)Bourveau, De~George, Ellahie, and Macciocchi}]{bourveau2022role}
Thomas Bourveau, Emmanuel~T De~George, Atif Ellahie, and Daniele Macciocchi. 2022.
\newblock The role of disclosure and information intermediaries in an unregulated capital market: Evidence from initial coin offerings.
\newblock \emph{Journal of Accounting Research}, 60(1):129--167.

\bibitem[{Chalkidis et~al.(2020)Chalkidis, Fergadiotis, Malakasiotis, Aletras, and Androutsopoulos}]{chalkidis-etal-2020-legal}
Ilias Chalkidis, Manos Fergadiotis, Prodromos Malakasiotis, Nikolaos Aletras, and Ion Androutsopoulos. 2020.
\newblock \href {https://doi.org/10.18653/v1/2020.findings-emnlp.261} {{LEGAL}-{BERT}: The muppets straight out of law school}.
\newblock In \emph{Findings of the Association for Computational Linguistics: EMNLP 2020}, pages 2898--2904, Online. Association for Computational Linguistics.

\bibitem[{Chalkidis et~al.(2023)Chalkidis, Garneau, Goanta, Katz, and S{\o}gaard}]{chalkidis-etal-2023-lexfiles}
Ilias Chalkidis, Nicolas Garneau, Catalina Goanta, Daniel Katz, and Anders S{\o}gaard. 2023.
\newblock \href {https://aclanthology.org/2023.acl-long.865} {{L}e{XF}iles and {L}egal{LAMA}: Facilitating {E}nglish multinational legal language model development}.
\newblock In \emph{Proceedings of the 61st Annual Meeting of the Association for Computational Linguistics (Volume 1: Long Papers)}, pages 15513--15535, Toronto, Canada. Association for Computational Linguistics.

\bibitem[{Chan et~al.(2020)Chan, Schweter, and M{\"o}ller}]{chan-etal-2020-germans}
Branden Chan, Stefan Schweter, and Timo M{\"o}ller. 2020.
\newblock \href {https://doi.org/10.18653/v1/2020.coling-main.598} {{G}erman{'}s next language model}.
\newblock In \emph{Proceedings of the 28th International Conference on Computational Linguistics}, pages 6788--6796, Barcelona, Spain (Online). International Committee on Computational Linguistics.

\bibitem[{Chen(2019)}]{chen2019information}
Kun Chen. 2019.
\newblock Information asymmetry in initial coin offerings (icos): Investigating the effects of multiple channel signals.
\newblock \emph{Electronic Commerce Research and Applications}, 36:100858.

\bibitem[{Devlin et~al.(2019)Devlin, Chang, Lee, and Toutanova}]{DBLP:conf/naacl/DevlinCLT19}
Jacob Devlin, Ming{-}Wei Chang, Kenton Lee, and Kristina Toutanova. 2019.
\newblock \href {https://doi.org/10.18653/v1/n19-1423} {{BERT:} pre-training of deep bidirectional transformers for language understanding}.
\newblock In \emph{Proceedings of the 2019 Conference of the North American Chapter of the Association for Computational Linguistics: Human Language Technologies, {NAACL-HLT} 2019, Minneapolis, MN, USA, June 2-7, 2019, Volume 1 (Long and Short Papers)}, pages 4171--4186. Association for Computational Linguistics.

\bibitem[{Fisch(2019)}]{Fisch2019}
Christian Fisch. 2019.
\newblock \href {https://doi.org/10.1016/j.jbusvent.2018.0} {{Initial coin offerings (ICOs) to finance new ventures}}.
\newblock \emph{Journal of Business Venturing}, 34(1):1--22.

\bibitem[{Florysiak(2022)}]{florysiak2022utility}
David Florysiak. 2022.
\newblock Utility tokens, markets in crypto assets regulation (micar), and the costs of being public.
\newblock \emph{Markets in Crypto Assets Regulation (MiCAR), and the Costs of Being Public (December 7, 2022)}.

\bibitem[{Florysiak and Schandlbauer(2022)}]{expertsorcharlatans}
David Florysiak and Alexander Schandlbauer. 2022.
\newblock \href {https://doi.org/10.1016/j.jbankfin.2022.1} {{Experts or charlatans? ICO analysts and white paper informativeness}}.
\newblock \emph{Journal of Banking \& Finance}, 139(C).

\bibitem[{Francis et~al.(2019)Francis, Van~Landeghem, and Moens}]{francis2019transfer}
Sumam Francis, Jordy Van~Landeghem, and Marie-Francine Moens. 2019.
\newblock Transfer learning for named entity recognition in financial and biomedical documents.
\newblock \emph{Information}, 10(8):248.

\bibitem[{Guha et~al.(2023)Guha, Nyarko, Ho, Ré, Chilton, Narayana, Chohlas-Wood, Peters, Waldon, Rockmore, Zambrano, Talisman, Hoque, Surani, Fagan, Sarfaty, Dickinson, Porat, Hegland, Wu, Nudell, Niklaus, Nay, Choi, Tobia, Hagan, Ma, Livermore, Rasumov-Rahe, Holzenberger, Kolt, Henderson, Rehaag, Goel, Gao, Williams, Gandhi, Zur, Iyer, and Li}]{guha2023legalbench}
Neel Guha, Julian Nyarko, Daniel~E. Ho, Christopher Ré, Adam Chilton, Aditya Narayana, Alex Chohlas-Wood, Austin Peters, Brandon Waldon, Daniel~N. Rockmore, Diego Zambrano, Dmitry Talisman, Enam Hoque, Faiz Surani, Frank Fagan, Galit Sarfaty, Gregory~M. Dickinson, Haggai Porat, Jason Hegland, Jessica Wu, Joe Nudell, Joel Niklaus, John Nay, Jonathan~H. Choi, Kevin Tobia, Margaret Hagan, Megan Ma, Michael Livermore, Nikon Rasumov-Rahe, Nils Holzenberger, Noam Kolt, Peter Henderson, Sean Rehaag, Sharad Goel, Shang Gao, Spencer Williams, Sunny Gandhi, Tom Zur, Varun Iyer, and Zehua Li. 2023.
\newblock \href {http://arxiv.org/abs/2308.11462} {Legalbench: A collaboratively built benchmark for measuring legal reasoning in large language models}.

\bibitem[{Hanley and Hoberg(2010)}]{hanley2010information}
Kathleen~Weiss Hanley and Gerard Hoberg. 2010.
\newblock The information content of ipo prospectuses.
\newblock \emph{The Review of Financial Studies}, 23(7):2821--2864.

\bibitem[{Henderson* et~al.(2022)Henderson*, Krass*, Zheng, Guha, Manning, Jurafsky, and Ho}]{hendersonkrass2022pileoflaw}
Peter Henderson*, Mark~S. Krass*, Lucia Zheng, Neel Guha, Christopher~D. Manning, Dan Jurafsky, and Daniel~E. Ho. 2022.
\newblock \href {https://arxiv.org/abs/2207.00220} {Pile of law: Learning responsible data filtering from the law and a 256gb open-source legal dataset}.

\bibitem[{Howell et~al.(2020)Howell, Niessner, and Yermack}]{howell2020initial}
Sabrina~T Howell, Marina Niessner, and David Yermack. 2020.
\newblock Initial coin offerings: Financing growth with cryptocurrency token sales.
\newblock \emph{The Review of Financial Studies}, 33(9):3925--3974.

\bibitem[{Hu and Liu(2004)}]{hu2004mining}
Minqing Hu and Bing Liu. 2004.
\newblock Mining opinion features in customer reviews.
\newblock In \emph{AAAI}, volume~4, pages 755--760.

\bibitem[{Karimov and W{\'o}jcik(2021)}]{karimov2021identification}
Bedil Karimov and Piotr W{\'o}jcik. 2021.
\newblock Identification of scams in initial coin offerings with machine learning.
\newblock \emph{Frontiers in Artificial Intelligence}, 4:718450.

\bibitem[{Keymanesh et~al.(2021)Keymanesh, Elsner, and Parthasarathy}]{keymanesh2021privacy}
Moniba Keymanesh, Micha Elsner, and Srinivasan Parthasarathy. 2021.
\newblock Privacy policy question answering assistant: A query-guided extractive summarization approach.
\newblock \emph{arXiv preprint arXiv:2109.14638}.

\bibitem[{Li et~al.(2018)Li, Wei, Liu, Yao, Chen, Yu, and Zhu}]{li2018incorporating}
Yiming Li, Baogang Wei, Yonghuai Liu, Liang Yao, Hui Chen, Jifang Yu, and Wenhao Zhu. 2018.
\newblock Incorporating knowledge into neural network for text representation.
\newblock \emph{Expert Systems with Applications}, 96:103--114.

\bibitem[{Liebau and Schueffel(2019)}]{liebau2019crypto}
Daniel Liebau and Patrick Schueffel. 2019.
\newblock Crypto-currencies and icos: Are they scams? an empirical study.
\newblock \emph{An Empirical Study (January 23, 2019)}.

\bibitem[{Liu et~al.(2021)Liu, Sheng, and Wang}]{liu2021technology}
Yukun Liu, Jinfei Sheng, and Wanyi Wang. 2021.
\newblock Technology and cryptocurrency valuation: Evidence from machine learning.
\newblock In \emph{Technology and Cryptocurrency Valuation: Evidence from Machine Learning: Liu, Yukun ajdnSheng, Jinfei ajd Wang, Wanyi}. SSRN.

\bibitem[{Mamakas et~al.(2022)Mamakas, Tsotsi, Androutsopoulos, and Chalkidis}]{mamakas-etal-2022-processing}
Dimitris Mamakas, Petros Tsotsi, Ion Androutsopoulos, and Ilias Chalkidis. 2022.
\newblock \href {https://doi.org/10.18653/v1/2022.nllp-1.11} {Processing long legal documents with pre-trained transformers: Modding {L}egal{BERT} and longformer}.
\newblock In \emph{Proceedings of the Natural Legal Language Processing Workshop 2022}, pages 130--142, Abu Dhabi, United Arab Emirates (Hybrid). Association for Computational Linguistics.

\bibitem[{Meoli and Vismara(2022)}]{meoli2022machine}
Michele Meoli and Silvio Vismara. 2022.
\newblock Machine-learning forecasting of successful icos.
\newblock \emph{Journal of Economics and Business}, 121:106071.

\bibitem[{Mikolov et~al.(2013)Mikolov, Sutskever, Chen, Corrado, and Dean}]{mikolov2013distributed}
Tomas Mikolov, Ilya Sutskever, Kai Chen, Greg~S Corrado, and Jeff Dean. 2013.
\newblock Distributed representations of words and phrases and their compositionality.
\newblock \emph{Advances in neural information processing systems}, 26.

\bibitem[{Momtaz(2021)}]{momtaz2021entrepreneurial}
Paul~P Momtaz. 2021.
\newblock Entrepreneurial finance and moral hazard: evidence from token offerings.
\newblock \emph{Journal of Business Venturing}, 36(5):106001.

\bibitem[{Morin et~al.(2021)Morin, Vasek, and Moore}]{morin2021detecting}
Andrew Morin, Marie Vasek, and Tyler Moore. 2021.
\newblock Detecting text reuse in cryptocurrency whitepapers.
\newblock In \emph{2021 IEEE International Conference on Blockchain and Cryptocurrency (ICBC)}, pages 1--5. IEEE.

\bibitem[{Niklaus et~al.(2023)Niklaus, Matoshi, Stürmer, Chalkidis, and Ho}]{niklaus2023multilegalpile}
Joel Niklaus, Veton Matoshi, Matthias Stürmer, Ilias Chalkidis, and Daniel~E. Ho. 2023.
\newblock \href {http://arxiv.org/abs/2306.02069} {Multilegalpile: A 689gb multilingual legal corpus}.

\bibitem[{Pont et~al.(2023)Pont, Galli, Loreggia, Pisano, Rovatti, and Sartor}]{pont2023legal}
Thiago~Dal Pont, Federico Galli, Andrea Loreggia, Giuseppe Pisano, Riccardo Rovatti, and Giovanni Sartor. 2023.
\newblock Legal summarisation through llms: The prodigit project.
\newblock \emph{arXiv preprint arXiv:2308.04416}.

\bibitem[{Ravichander et~al.(2019)Ravichander, Black, Wilson, Norton, and Sadeh}]{ravichander-etal-2019-question}
Abhilasha Ravichander, Alan~W Black, Shomir Wilson, Thomas Norton, and Norman Sadeh. 2019.
\newblock \href {https://doi.org/10.18653/v1/D19-1500} {Question answering for privacy policies: Combining computational and legal perspectives}.
\newblock In \emph{Proceedings of the 2019 Conference on Empirical Methods in Natural Language Processing and the 9th International Joint Conference on Natural Language Processing (EMNLP-IJCNLP)}, pages 4947--4958, Hong Kong, China. Association for Computational Linguistics.

\bibitem[{Salinas~Alvarado et~al.(2015)Salinas~Alvarado, Verspoor, and Baldwin}]{salinas-alvarado-etal-2015-domain}
Julio~Cesar Salinas~Alvarado, Karin Verspoor, and Timothy Baldwin. 2015.
\newblock \href {https://aclanthology.org/U15-1010} {Domain adaption of named entity recognition to support credit risk assessment}.
\newblock In \emph{Proceedings of the Australasian Language Technology Association Workshop 2015}, pages 84--90, Parramatta, Australia.

\bibitem[{Samieifar and Baur(2021)}]{samieifar2021read}
Shadi Samieifar and Dirk~G Baur. 2021.
\newblock Read me if you can! an analysis of ico white papers.
\newblock \emph{Finance Research Letters}, 38:101427.

\bibitem[{Sharpe and Decker(2022)}]{sharpe-decker-2022-prospectus}
Jared Sharpe and Keith Decker. 2022.
\newblock \href {https://doi.org/10.18653/v1/2022.finnlp-1.21} {Prospectus language and {IPO} performance}.
\newblock In \emph{Proceedings of the Fourth Workshop on Financial Technology and Natural Language Processing (FinNLP)}, pages 154--162, Abu Dhabi, United Arab Emirates (Hybrid). Association for Computational Linguistics.

\bibitem[{Sm{\u{a}}du et~al.(2022)Sm{\u{a}}du, Dinic{\u{a}}, Avram, Cercel, Pop, and Cercel}]{smadu-etal-2022-legal}
R{\u{a}}zvan-Alexandru Sm{\u{a}}du, Ion-Robert Dinic{\u{a}}, Andrei-Marius Avram, Dumitru-Clementin Cercel, Florin Pop, and Mihaela-Claudia Cercel. 2022.
\newblock \href {https://doi.org/10.18653/v1/2022.nllp-1.29} {Legal named entity recognition with multi-task domain adaptation}.
\newblock In \emph{Proceedings of the Natural Legal Language Processing Workshop 2022}, pages 305--321, Abu Dhabi, United Arab Emirates (Hybrid). Association for Computational Linguistics.

\bibitem[{Stiennon et~al.(2020)Stiennon, Ouyang, Wu, Ziegler, Lowe, Voss, Radford, Amodei, and Christiano}]{stiennon2020learning}
Nisan Stiennon, Long Ouyang, Jeffrey Wu, Daniel Ziegler, Ryan Lowe, Chelsea Voss, Alec Radford, Dario Amodei, and Paul~F Christiano. 2020.
\newblock Learning to summarize with human feedback.
\newblock \emph{Advances in Neural Information Processing Systems}, 33:3008--3021.

\bibitem[{Tao et~al.(2018)Tao, Deokar, and Deshmukh}]{tao2018analysing}
Jie Tao, Amit~V Deokar, and Ashutosh Deshmukh. 2018.
\newblock Analysing forward-looking statements in initial public offering prospectuses: a text analytics approach.
\newblock \emph{Journal of Business Analytics}, 1(1):54--70.

\bibitem[{Thewissen et~al.(2022)Thewissen, Shrestha, Torsin, and Pastwa}]{unpackingblackbox}
James Thewissen, Prabal Shrestha, Wouter Torsin, and Anna~M Pastwa. 2022.
\newblock Unpacking the black box of ico white papers: A topic modeling approach.
\newblock \emph{Journal of Corporate Finance}, 75:102225.

\bibitem[{Thewissen et~al.(2023)Thewissen, Thewissen, Torsin, and Arslan-Ayaydin}]{thewissen2023linguistic}
Jennifer Thewissen, James Thewissen, Wouter Torsin, and {\"O}zg{\"u}r Arslan-Ayaydin. 2023.
\newblock Linguistic errors and investment decisions: the case of ico white papers.
\newblock \emph{The European Journal of Finance}, 29(7):826--868.

\bibitem[{Trautmann et~al.(2022)Trautmann, Petrova, and Schilder}]{trautmann2022legal}
Dietrich Trautmann, Alina Petrova, and Frank Schilder. 2022.
\newblock Legal prompt engineering for multilingual legal judgement prediction.
\newblock \emph{arXiv preprint arXiv:2212.02199}.

\bibitem[{Vaswani et~al.(2017)Vaswani, Shazeer, Parmar, Uszkoreit, Jones, Gomez, Kaiser, and Polosukhin}]{vaswani2017attention}
Ashish Vaswani, Noam Shazeer, Niki Parmar, Jakob Uszkoreit, Llion Jones, Aidan~N Gomez, {\L}ukasz Kaiser, and Illia Polosukhin. 2017.
\newblock Attention is all you need.
\newblock \emph{Advances in neural information processing systems}, 30.

\bibitem[{Yang et~al.(2023)Yang, Liu, and Wang}]{2306.06031}
Hongyang Yang, Xiao-Yang Liu, and Christina~Dan Wang. 2023.
\newblock \href {http://arxiv.org/abs/arXiv:2306.06031} {Fingpt: Open-source financial large language models}.

\bibitem[{Yen et~al.(2021)Yen, Wang, and Chen}]{yen2021different}
Ju-Chun Yen, Tawei Wang, and Yu-Hung Chen. 2021.
\newblock Different is better: how unique initial coin offering language in white papers enhances success.
\newblock \emph{Accounting \& Finance}, 61(4):5309--5340.

\bibitem[{Yu et~al.(2023)Yu, Quartey, and Schilder}]{yu-etal-2023-exploring}
Fangyi Yu, Lee Quartey, and Frank Schilder. 2023.
\newblock \href {https://doi.org/10.18653/v1/2023.findings-acl.858} {Exploring the effectiveness of prompt engineering for legal reasoning tasks}.
\newblock In \emph{Findings of the Association for Computational Linguistics: ACL 2023}, pages 13582--13596, Toronto, Canada. Association for Computational Linguistics.

\bibitem[{Zhang et~al.(2019)Zhang, Aerts, Lu, and Pan}]{zhang2019readability}
Shuyu Zhang, Walter Aerts, Liping Lu, and Huifeng Pan. 2019.
\newblock Readability of token whitepaper and ico first-day return.
\newblock \emph{Economics Letters}, 180:58--61.

\bibitem[{Zhong et~al.(2020)Zhong, Xiao, Tu, Zhang, Liu, and Sun}]{zhong-etal-2020-nlp}
Haoxi Zhong, Chaojun Xiao, Cunchao Tu, Tianyang Zhang, Zhiyuan Liu, and Maosong Sun. 2020.
\newblock \href {https://doi.org/10.18653/v1/2020.acl-main.466} {How does {NLP} benefit legal system: A summary of legal artificial intelligence}.
\newblock In \emph{Proceedings of the 58th Annual Meeting of the Association for Computational Linguistics}, pages 5218--5230, Online. Association for Computational Linguistics.

\bibitem[{Zufall et~al.(2020)Zufall, Hamacher, Kloppenborg, and Zesch}]{https://doi.org/10.48550/arxiv.2004.03422}
Frederike Zufall, Marius Hamacher, Katharina Kloppenborg, and Torsten Zesch. 2020.
\newblock \href {https://doi.org/10.48550/ARXIV.2004.03422} {A legal approach to hate speech: Operationalizing the eu's legal framework against the expression of hatred as an nlp task}.

\end{thebibliography}
\onecolumn
\appendix

\section{MiCAR Content Requirements Categories}
\label{sec:appendixA}
\subsection{Categories for Crypto Assets ``other than''}
Found in Annex I of the regulation.
\begin{itemize}
\item \textbf{Part A: }Information about the offeror or the person seeking admission to trading
\item \textbf{Part B:} Information about the issuer, if different from the offeror or person seeking admission to trading
\item \textbf{Part C:} Information about the operator of the trading platform in cases where it draws up the crypto asset white paper
\item \textbf{Part D:} Information about the crypto asset project
\item \textbf{Part E:} Information about the offer to the public of crypto assets or their admission to trading
\item \textbf{Part F:} Information about the crypto assets
\item \textbf{Part G:} Information on the rights and obligations attached to the crypto assets
\item \textbf{Part H:} Information on the underlying technology
\item \textbf{Part I:} Information on the risks
\end{itemize}

\begin{table*}[h]
\ra{1.25}
\centering
\begin{tabularx}{\textwidth}{@{}llllllllll@{}}

\toprule
                & Part A & Part B & Part C & Part D & Part E & Part F & Part G & Part H & Part I\\ 
\hline
Crypto ``other than'' &   10   &   8    &   9   &   6   &   19    &   2    &   10   &  5 & 5\\ \bottomrule
\end{tabularx}
\caption{Category count for each Part in the content requirements, for crypto assets ``other than''.}
\label{table:catcountsA}
\end{table*}
\subsection{Categories for Asset-Referenced Tokens and E-money Tokens }
Found in Annexes II and III of the regulation.
\begin{itemize}
    \item \textbf{Part A:} Information about the issuer of the token
    \item \textbf{Part B:} Information about the token
    \item \textbf{Part C:} Information about the offer to the public of the token or its admission to trading
    \item \textbf{Part D:} Information on the rights and obligations attached to the token
    \item \textbf{Part E:} Information on the underlying technology
    \item \textbf{Part F:} Information on the risks
    \item \textbf{Part G:} Information on the reserve of assets
\end{itemize}
\begin{table*}[h]
\ra{1.25}
\centering
\begin{tabularx}{\textwidth}{@{}lccccccc@{}}

\toprule
                & Part A & Part B & Part C & Part D & Part E & Part F & Part G \\ 
\hline
Asset-referenced Tokens &   12   &   5    &   15   &   17   &   5    &   6    &   5    \\
E-money tokens          &   13   &   3    &    4   &    8   &   6    &   3    &   -    \\ \bottomrule
\end{tabularx}
\caption{Category count for each Part in the content requirements, for asset-referenced and e-money tokens.}
\label{table:catcountsB}
\end{table*}

\end{document}